\documentstyle[aps,epsf,prl,multicol]{revtex}

\begin{document}
\title{Transport properties of 1D disordered models: a novel approach}
\author{V. Dossetti-Romero, F.M. Izrailev and A.A. Krokhin}
\address{
Instituto de F\'isica, Universidad Aut\'onoma de Puebla, Apdo.
Postal J-48, Puebla 72570, Mexico\\}

\date{\today}
\maketitle
\begin{abstract}
A new method is developed for the study of transport properties of
1D models with random potentials. It is based on an exact
transformation that reduces discrete Schr\"odinger equation in the
tight-binding model to a two-dimensional Hamiltonian map. This map
describes the behavior of a classical linear oscillator under
random parametric delta-kicks. We are interested in the
statistical properties of the transmission coefficient $T_L$ of a
disordered sample of length $L$. In the ballistic regime we derive
expressions for the mean value of the transmission coefficient
$T_L$, its second moment and variance, that are more accurate than
the existing ones. In the localized regime we analyze the global
characteristics of $\ln T_L$, and demonstrate that its
distribution function approaches the Gaussian form if
$L\rightarrow \infty$. For any finite $L$ there are deviations
from the Gaussian law that originate from the subtle correlation
effects between different trajectories of the Hamiltonian map.
\end{abstract}

\begin{multicols}{2}

\section{Introduction}
In recent years the interest to 1D models with random potentials
has been significantly increased. There are two reasons for this.
First, it is expected that 1D systems may elucidate the origin of
the famous {\it single parameter scaling} (SPS)\cite{AALR79} for
transport characteristics of disordered conductors. As was claimed
in Refs.\cite{alt}, a random character of the fluctuations of the
Lyapunov exponent for finite-length samples, that was originally
used\cite{ATAF80} to justify SPS, is not correct. Specifically, it
was shown that in the vicinity of the band edges the SPS does not
hold \cite{alt}. This result is important not only from the
theoretical viewpoint, but also for the experiment (see discussion
and references in \cite{alt}).

Another reason of the growing interest to 1D random models is due
to recent results on the {\it correlated disorder}. In early
studies of the 1D tight-binding model with a specific site
potential (the so-called random dimer model \cite{dimer}), it was
demonstrated a highly non-trivial role of {\it short-range}
correlations. It was found that for discrete values of energy $E$
that are determined by the model parameters, a random dimer turns
out to be fully transparent (for other examples, see
\cite{DR79}). Practically, this leads to an emergence of a finite range
of energies close to the resonant one, where the localization
length $l_\infty$ of eigenstates is larger than the size $L$ of
finite samples. Since the number of such states is of the order of
$\sqrt L$, this effect was claimed to have practical importance in
application to polymer chains.

Although the localization length $l_\infty$ diverges at discrete
values of $E$, this model does not exhibit a mobility edge.
However, the situation was found to be very different for the case
of {\it long-range} correlations \cite{lyra}. As was shown in
Ref.\cite{IK99}, one can construct such random (correlated)
potentials that result (for a weak disorder) in a band of a
complete transparency.  The position and the width of the window
of transparency can be controlled by the form of the binary
correlator of the weak random potential \cite{filter}.

The role of long-range correlations has been studied in details
for the tight-binding Anderson-type model \cite{IK99}, and for the
Kronig-Penney model with randomly distributed amplitudes
\cite{KI00} and positions of delta-peaks \cite{IKU01}. The results
have been also extended to a single-mode waveguide with random
surface profiles \cite{IM01}. The prediction of the theory
\cite{IK99} has been verified experimentally \cite{KIKS00},
when studying the transport properties of a single-mode
electromagnetic waveguide with point-like scatterers. The latter
have been intentionally inserted into the waveguide in a way to
provide a random potential with slowly decaying binary correlator.
Very recently \cite{IM02} the existence of the mobility edges was
predicted for waveguides with a finite number of propagating
channels (quasi-1D system) with long-range correlations in random
surface scattering potential.

An appropriate tool to study the correlated disorder in 1D
Anderson and Kronig-Penney models is the so-called Hamiltonian map
(HM) approach. The key point of this approach is a transformation
that reduces a discrete 1D Schr\"odinger equation to the classical
two-dimensional Hamiltonian map. The properties of trajectories of
this map are related to transport properties of a quantum model.
The geometrical aspects of the HM approach turn out to be helpful
in qualitative analysis as well as in deriving analytical
formulas.

Originally, the HM approach was proposed in Ref.\cite{IKT95} in
order to study extended states in the dimer model. It was
demonstrated that with the help of this approach it is easy to
understand the mechanism for the emergence of the extended states.
Specifically, the conditions for resonant energies can be easily
obtained not only for dimers, but also for $N-$mers (when blocks
of the length $N$ with the same site energy $\epsilon$ appear
randomly in a site potential). More important is that the
expression for the localization length $l_\infty$ in the vicinity
of the band edges has been derived in a more general form, as
compared to that found in earlier studies.

The generalization of the HM approach to correlated 1D-potentials
\cite{IK99,KI00,IKU01} has led to an understanding of the role
of long-range correlations for the emergence of extended quantum
states in random potentials. As a result, a new principle was
proposed for filtering of stochastic and digital signals using
random systems with correlated disorder \cite{filter}.

So far, the HM approach was used for calculations of the
localization length in infinite samples. A more serious problem
arises when studying the transport properties of finite samples of
length $L$. It is assumed \cite{ATAF80} that statistical
properties of the transmission coefficient $T_L$ are entirely
determined by the finite-length Lyapunov exponent (FLLE) which
fluctuates with different realizations of the random potential.
According to this point, many studies of solid state models with
random potentials are directly related to the analysis of
statistical properties of the FLLE.

In this paper we propose a new method based on the HM approach,
that allows to obtain main transport characteristics in 1D random
models. To illustrate this approach, we take the standard Anderson
model with weak white-noise potential, and consider two limit
cases of the ballistic and localized transport. We show that in
this way one can relatively easily derive some of known results,
as well as obtain the new ones.

\section{The Hamiltonian map approach}

It is well known that the discrete 1D Anderson model can be
written in the following form of Schr\"odinger equation for
stationary eigenstates $\psi_{n}$,
\begin{equation}
\psi_{n+1}+\psi_{n-1}=\left( E+\epsilon_{n} \right)\psi_{n}.
\label{eq:1}
\end{equation}
\nolinebreak
Here $E$ is the energy of a specific eigenstate, and
$\epsilon_{n}$ is the potential energy at site $n$. It is
convenient to represent Eq.(\ref{eq:1}) in the form of
two-dimensional Hamiltonian map
\cite{IKT95,IRT98},
\begin{equation}
\left( \begin{array}{c}
             x_{n+1} \\ p_{n+1}
        \end{array} \right) =
\left( \begin{array}{clcr}
             \cos \mu + A_{n} \sin \mu &\,\,\,\,   \sin \mu \\
             A_{n} \cos \mu - \sin \mu & \,\,\,\,  \cos \mu
        \end{array} \right)
\left( \begin{array}{c}
             x_{n} \\ p_{n}
        \end{array} \right).
\label{eq:2}
\end{equation}
\nolinebreak
In this map the canonical variables $x_n=\psi_n$ and $p_n=(\psi_n
\cos\mu-\psi_{n-1})/\sin \mu$ correspond to the position and
momentum of a linear oscillator subjected to linear time-periodic
delta-kicks. The amplitudes $A_n$ of the kicks are proportional to
the site potential in Eq.(\ref{eq:1}),
$A_{n}=-\epsilon_{n}/\sin\mu$, and the parameter $\mu$ is
determined by the energy of an eigenstate, $2 \cos\mu = E$.

One can see that the representation (\ref{eq:2}) is the
Hamiltonian version of the standard transfer matrix method.
Indeed, starting from initial values $\psi_0$ and $\psi_{-1}$, one
can compute $\psi_n$ and $\psi_{n-1}$ according to Eq.(\ref{eq:1})
or Eq.(\ref{eq:2}). The Lyapunov exponent $\Lambda$ and,
therefore, the localization length $\l_{\infty}=
\Lambda^{-1}$ is obtained in the limit $n\rightarrow\infty$ (see
below). It turns out that the Hamiltonian representation
(\ref{eq:2}) is more convenient than the standard one based on
Eq.(\ref{eq:1}). This is due a possibility to introduce the
classical phase space in order to study the properties of a
trajectory $\{x_n, p_n\}$.

Mathematically, the Anderson localization corresponds to the
parametric instability of a linear oscillator associated with the
map (\ref{eq:2}), see details in \cite{Luca} and some applications
in \cite{B02}. One should stress that the unbounded classical
trajectories of the map (\ref{eq:2}) do not correspond to the
eigenstates of Eq.(\ref{eq:1}), however, they give a correct value
of the localizaion length through the Lyapunov exponent. On the
contrary, if an eigenstate $\psi_n$ of Eq.(\ref{eq:1}) is {\it
extended} (delocalized) in the infinite configuration space
$\{n\}$, the corresponding trajectory of the map (\ref{eq:2}) is
{\it bounded} in the phase space $\{x_n, p_n\}$. In this case the
trajectory specified by the value $\mu$ has a direct
correspondence to the eigenstate with energy $E$. Therefore, the
structure of such eigenstates can be studied by analyzing the
properties of the trajectories in the phase space.

For our analysis it is convenient to represent the map of Eq.\
(\ref{eq:2}) in the action-angle variables
$\left(r,\theta\right)$. Using the standard transformation
$x=r\sin\theta$ and $p=r\cos\theta$, the map can be rewritten as
follows,
\begin{equation}
\begin{array}{l}
r_{n+1} = r_{n}D_{n}, \nonumber \\ \\
\sin\theta_{n+1} = D_{n}^{-1}
\left[ \sin\left(\theta_{n}-\mu\right)
- A_{n}\sin\theta_{n}\sin\mu \right], \\ \\
\cos\theta_{n+1} = D_{n}^{-1}
\left[ \cos\left(\theta_{n}-\mu\right)
+ A_{n}\sin\theta_{n}\cos\mu \right] \nonumber
\end{array}
\label{eq:3}
\end{equation}
\nolinebreak
where
\begin{equation}
D_{n} = \sqrt{1 + A_{n}\sin 2\theta_{n} +
A_{n}^{2}\sin^{2}\theta_{n}} .
\label{eq:4}
\end{equation}

Eqs.(\ref{eq:3}) and (\ref{eq:4}) allow one to represent the
Lyapunov exponent $\Lambda$ in terms of $A_n$ and $\theta_n$ (for
energies not close to the band edges, see details in
\cite{IRT98}),
\begin{equation}
\begin{array}{ll}
\Lambda \equiv l_{\infty}^{-1}  =
\lim\limits_{L \rightarrow \infty} \frac{1}{L} \sum^{L}\limits_{n=1}
\ln \left( \frac{r_{n+1}}{r_n}\right)
\\ ~~~~~~~~~\\
= \frac{1}{2} \langle \ln \left( 1 + A_{n}\sin 2\theta_{n} +
A_{n}^{2}\sin^{2}\theta_{n}
\right) \rangle_n.
\label{lyap}
\end{array}
\end{equation}
Here the angle brackets $\langle ... \rangle_n $ stand for the
average along a trajectory (over $n$).

The relation (\ref{lyap}) is valid for {\it arbitrary} potential
$\epsilon_n$, no matter, weak or strong, random or deterministic,
provided that the parameter $\mu$ corresponds to the value of $E$
taken {\it inside} the energy spectrum. Note that for a {\it weak}
potential the spectrum remains unperturbed ($\mid E\mid < 2 $) in
the lowest (Born) approximation.

From Eqs.(\ref{eq:3}) and (\ref{eq:4}) it follows that actually
the small parameter is $A_n=-\epsilon_n/\sin \mu \ll 1$, rather
than $\mid\epsilon_n\mid \ll 1$. Therefore, close to the edges of
the energy band where $|\sin \mu|\approx 0$, the standard
perturbation theory fails. That is why for energies close to the
band edges one needs to use specific methods for calculation of
the localization length (see \cite{IRT98} and references therein).

If the energy is not close to the band edges, $\mid E \mid =2$, or
to the band center, $E=0$, the standard perturbation theory with
respect to $A_n$ is applicable. For a weak  {\it uncorrelated}
potential ("white noise") the distribution of $\theta_n$ is
homogeneous, ${\cal P} (\theta) = 1/2\pi$, i.e., the phases
$\theta_n$ are independent of the potential $\epsilon_n$.
Therefore, instead of the average along the trajectory, one can
perform an ensemble average over $\epsilon_n$ and $\theta_n$
independently. The result for $l_\infty$ can be obtained easily,
by keeping the linear and quadratic terms in the expansion of the
logarithm in Eq.(\ref{lyap}),
\begin{equation}
l_{\infty}^{-1} = \frac{\left< \epsilon^{2}
\right>}{8\sin^{2}\mu} = \frac{\left<\epsilon^{2}\right>}
{8\left( 1-\frac{E^2}{4} \right)}.
\label{eq:5}
\end{equation}
Here and below the brackets $\left<...\right>$ stand for the
average over disorder.

For the first time the result (\ref{eq:5}) was obtained by
Thouless \cite{Thouless} by another method. It is interesting to
note that at the center of the energy band, $E=0$, the correct
expression for the localization length is slightly different from
Eq.(\ref{eq:5}). At this point the Born approximation is invalid
since the kinetic energy vanishes.

Using the HM approach it is easy to see that at $E=0$ (where
$\mu=\pi/2$), classical trajectories reveal a mixture of the
periodic rotation with period $4$ (in number of kicks) around the
origin $p=x=0$, and a very slow diffusion in $\theta$ and $r$. As
a result, the distribution function $\cal P(\theta)$ turns out to
be slightly modulated over $\theta$ by a periodic function with
period $\pi/2$ (see details and discussion in
\cite{IRT98}). This leads to an anomalous contribution of the
fourth Fourier harmonic, which needs to be taken into account in
addition to the contribution of the zero harmonic.

It is important to note that the scaling hypothesis is known to be
valid at the band center. On the other hand, the phases $\theta_n$
are not distributed randomly in this case. Therefore, for energies
close to $E=0$ the random phase approximation which is often
assumed to be a core of the SPS conjecture, is not valid. This
fact supports the statement of Refs.\cite{alt} that the origin of
the SPS is not in the randomness of phases.

A more complicated situation arises for energies close to the band
edges, $\delta = 2-|E| \ll 1$. However, even in this case the
non-perturbative study (with respect to $A_n$) of classical
trajectories of the map (\ref{eq:2}) allows to find the analytical
expression for the Lyapunov exponent $\Lambda(\delta)$
\cite{IRT98}.

The HM approach is applicable also for the analysis of the
transport properties of {\it finite} samples. The transmission
coefficient $T_L$ of a sample of size $L$ can be expressed in
terms of the radial variable as follows \cite{KTI97},
\begin{equation}
T_{L} = \frac{2}{1 + \frac{1}{2} \left( r_{L1}^{2} + r_{L2}^{2} \right)}.
\label{eq:6}
\end{equation}
Here $r_{L, i=1,2}^{2}$ are the radial coordinates of the last
points ($n=L$) of the two trajectories that start from $\left(
r_{0}^{(1)},\theta_{0}^{(1)} \right) = \left( 1,0 \right)$ and
\mbox{$\left( r_{0}^{(2)},\theta_{0}^{(2)}
\right)= \left( 1,\pi/2 \right)$} respectively. The values of
$r_{L1}^{2}$ and $r_{L2}^{2}$ are calculated numerically by
iterating the map (\ref{eq:3}). From equations (\ref{eq:3}) one
gets,
\begin{equation}
\label{radii}
r_{L1}^{2}=\prod_{n=0}^{L-1}\left( D_{n}^{(1)} \right)^2,
\,\,\,\,\,\,\,\,\,
r_{L2}^{2}=\prod_{n=0}^{L-1}\left( D_{n}^{(2)} \right)^2\,\,\,.
\end{equation}

In what follows we apply Eq.(\ref{eq:6}) for uncorrelated random
potential in two limit cases of ballistic and localized regimes.
We assume that the site energies $\epsilon_n$ are distributed
randomly and homogeneously within the interval $|\epsilon_n| <
W/2$. We consider the case of a weak disorder when the variance
$\left<\epsilon_n^2\right> = W^2/12$ is small,
$\left<\epsilon_n^2\right>\ll 1$. Therefore, the inverse
localization length is given by,
\begin{equation}\label{loc}
l_\infty^{-1} = \frac{W^2} {96\left( 1-\frac{E^2}{4}
\right)}
\end{equation}
for energies not very close to band edges and to the center of the
energy band.


\section{Ballistic regime.}

Let us start with the ballistic regime for which the size $L$ of a
sample is much less than the localization length $l_\infty$,
\begin{equation}\label{ballistic}
\lambda \equiv \frac{L}{l_\infty} \ll 1.
\end{equation}
In this case it is convenient to introduce a new parameter,
\begin{equation}
Z_{L} = \frac{r_{L1}^{2} + r_{L2}^{2}}{2},
\label{eq:8}
\end{equation}
which is close to unity since both radii $r_{L, i=1,2}$ increase
in "time" $L$ very slowly, $r_L \sim \exp(\lambda)$. Therefore,
the transmission coefficient
\begin{equation}
T_{L} = \frac{2}{1 + Z_{L}} = \frac{2}{1+\exp(\ln Z_L)}
\end{equation}
can be evaluated perturbatively in terms of $\ln Z_L$. One can see
that the statistical properties of the transmission coefficient
$T_L$ are entirely determined by the properties of $Z_L$. The
latter is related to the two trajectories of the classical map. In
the ballistic regime a trajectory of this map exhibits fast
rotation over angle $\theta$ and slow diffusion in radial
direction (or, the same, in energy of the classical oscillator).

In the first line we are interested in the mean value
$\left<T_L\right>$ of the transmission coefficient, and in its
second moment $\left<T_L^2\right>$. Keeping the terms up to
$O\left(\ln^2 Z_L\right)$ in the expansion of $T_L$, we obtain
that the quadratic term $O\left(\ln^2 Z_{L}\right)$ does not
contribute to $T_L$,
\begin{equation}\label{eq:27x}
T_{L} \approx 1- \frac{1}{2} \ln Z_L + O \left(
\ln^{3} Z_L \right) \; \; .
\end{equation}
Similarily, we get,
\begin{equation}
T_{L}^2 \approx 1 - \ln Z_L + \frac{1}{4}\ln^{2}Z_L + O
\left(\ln^3 Z_L \right) \; \; .
\label{eq:28x}
\end{equation}

After some straightforward calculations \cite{long} that involve
the map (\ref{eq:3}), the following expression for the mean value
of $\ln Z_L$ is obtained,
\begin{eqnarray}
\left< \ln Z_L \right> & \approx & 3\lambda
- \frac{1}{2} \left( \frac{1}{2}-\lambda \right) S_2 -
\lambda^2  - \frac{1}{8} S_4,
\label{eq:29x}
\end{eqnarray}
where
\begin{equation}\label{S2}
S_2 =
\left< \sum_{n=0}^{L-1} A_{n}^{2} \sin \left( 2\theta_{n}^{(1)}
\right)\sin \left( 2\theta_{n}^{(2)} \right) \right>,
\end{equation}
and
\begin{eqnarray}
S_4 =
\Biggl<
\sum_{n>k}^{L-1}
\sum_{k=0}^{L-1} A_{n}^{2} \, A_{k}^{2} \,
\sin \left( 2\theta_{n}^{(1)} \right)
\sin \left( 2\theta_{n}^{(2)} \right)
\nonumber \\
\times \, \sin \left( 2\theta_{k}^{(1)} \right)
\sin \left( 2\theta_{k}^{(2)} \right)
\Biggr>\,\,\,.
\label{S4}
\end{eqnarray}
Here the terms $S_2$ and $S_4$ describe the correlations between
the phases $\theta_{n}^{(1)}$ and $\theta_{n}^{(2)}$ of the two
classical trajectories that start from two complementary initial
conditions, see Section 2. The presence of these correlation terms
in Eq.(\ref{eq:29x}) is an important fact that strongly restricts
the analytical treatment.

Let us analyze the term $S_2$. Taking into account that the
fluctuations of $A_n$ and $\theta_n$ are statistically
independent, we get,
\begin{equation}\label{S2R2}
S_2 = 8 l_\infty^{-1}  \sum_{n=0}^{L-1} R_2(\lambda_n),
\end{equation}
where we introduced the two-point correlator $R_2$ which depends
on the scaling parameter $\lambda_n=n/l_{\infty}$,
\begin{equation}\label{R2}
R_2(\lambda_n) = \left<\sin \left( 2\theta_{n}^{(1)}
\right) \sin \left( 2\theta_{n}^{(2)} \right) \right>.
\end{equation}
Here the average is taken over the disorder for a {\it fixed}
number of kicks $n = 1,...,L$.

In Fig. 1 we show numerical data for the correlator $R_2$ for a
wide range of the parameter $\lambda_n$ that covers metallic,
$\lambda_n \ll 1$, and localized, $\lambda_n \gg 1$ regimes. In
average, the correlator $R_2$ changes from $-1/2$ for the
ballistic regime to $1/2$ for the localized regime. This graph
shows that the correlations give different contributions in the
ballistic and localized regimes.

\begin{figure}[!tb]
\begin{center}
\leavevmode
\epsfysize=7cm
\epsfbox{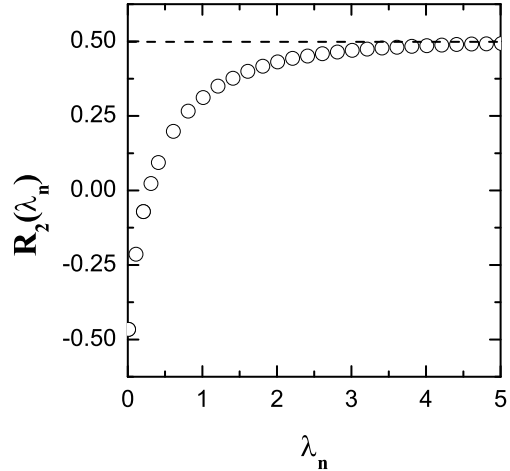}
\narrowtext
\caption{Numerical data for the correlator $R_2$ versus the
scaling parameter $\lambda_n = n/l_\infty $. The transition
from the ballistic to localized regime is shown for for $E=1.5$
and $W=0.1$. The average was done over $10^4$ realizations of the
disorder. An additional "window moving" average was performed in order
to reduce fluctuations.}
\label{fig:1}
\end{center}
\end{figure}

\vspace{-1.0cm}
\begin{figure}[!tb]
\begin{center}
\leavevmode
\epsfysize=7cm
\epsfbox{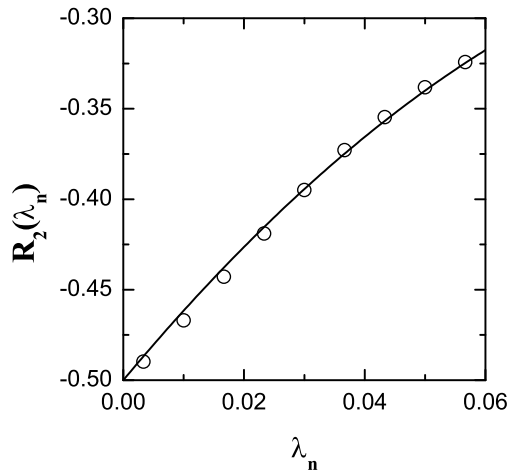}
\caption{Numerical data (open circles) for $R_2(\lambda_n)$
versus Eq.(\ref{eq:12}) (solid curve) for the same parameters as
in Fig.1. and $\lambda_n\ll 1$}
\label{fig:2}
\narrowtext
\end{center}
\end{figure}

In order to evaluate analytically the correlator $R_2$ in the
ballistic regime, we use the approximate map for the angle
$\theta_n$. It is obtained from Eq.(\ref{eq:3}) in the limit
$|\epsilon_n| \ll 1$,
\begin{equation}
\theta_{k+1} = \theta_{k} - \mu + \epsilon_{k} \frac{\sin^2
\theta_{k}}{\sin \mu}.
\label {eq:11}
\end{equation}
Using this recursion, one can express an angle $\theta_n$ in terms
of the amplitudes $\epsilon_0, \epsilon_, ... , \epsilon_{n-1}$ of
all the previous kicks for a fixed value of $\mu$. Then the
following expression for $R_2$ can be obtained \cite{long},
\begin{equation}
R_2(\lambda_n) = - \frac{1}{2} + 4 \lambda_n - 16 \lambda^2_n.
\label{eq:12}
\end{equation}
In Fig.2 we plot this formula together with numerical data for the
ballistic regime, $\lambda_n \ll 1$. One can see that there is a
complete agreement between numerical and analytical results.

The four-point correlator $S_4$, see Eq.(\ref{S4}), can be
calculated analytically in a similar way \cite{long},
\begin{equation}
\label{S4an}
S_4(\lambda) = 8 \lambda^2 + O \left( \lambda^3 \right) \; \; .
\end{equation}
Substitution of Eqs.(\ref{eq:12}) and (\ref{S4an}) into
Eq.(\ref{eq:29x}), gives the following formula for the mean value
of $\ln Z_L$
\begin{equation}
\left< \ln Z_L \right> \approx 4 \lambda - 8 \lambda^2
+ O \left( \lambda^3 \right) \; \; .
\label{eq:32x}
\end{equation}
In the same way we calculate the mean value of the second moment,
\begin{eqnarray}
\left< \mbox{ln}^2 Z_{L} \right> & \approx & 32 \lambda^2
+ O \left( \lambda^3 \right) \; \; .
\label{eq:34x}
\end{eqnarray}

Substituting Eqs.(\ref{eq:32x}) and (\ref{eq:34x}) into
Eqs.(\ref{eq:27x}) and (\ref{eq:28x}) respectively, we obtain,
\begin{equation}\label{mean}
\left< T_{L} \right> =  1 - 2 \lambda + 4 \lambda^2 + O
\left(\lambda^3 \right) \; \; ,
\end{equation}
and
\begin{equation}\label{square}
\left< T_{L}^2 \right> =  1 - 4 \lambda + 16 \lambda^2
+ O \left(\lambda^3 \right) \; \; .
\end{equation}
As a result, the variance reads as follows,
\begin{equation}\label{var}
\text{Var} (T_L) = \left< T_{L}^2 \right> - \left< T_{L} \right>^2 = 4
\lambda^2 + O
\left(\lambda^3 \right).
\end{equation}
The latter expression for $ \text{Var} (T_L) $ is known in the
literature (see, for example, \cite{Makarov}). However, from the
analysis of Eqs.(\ref {mean}) and (\ref{square}) one can obtain a
more accurate expression that depends on higher powers of
$\lambda$. Indeed, the expansions in (\ref{mean}) and
(\ref{square}) can be considered as asymptotics of the following
"exact" formulas,
\begin{equation}
\label{eq:35xb}
\left< T_{L} \right> =  \frac{1}{1 +2 \lambda}\,\,\,,
\end{equation}
and
\begin{equation}
\label{eq:36xb}
\left< T_{L}^2 \right>  =  \frac{1}{1+ 4 \lambda} \,\,\,.
\end{equation}
We have tested these expressions and found that they fit the
numerical data much better than Eqs.(\ref{mean}) and
(\ref{square}) \cite{long}. By combining Eq.(\ref{eq:35xb}) with
(\ref{eq:36xb}), one can obtain the following expression for the
variance of the transmission coefficient,
\begin{equation}
\text{Var} (T_L) = \frac{4\lambda^2}{(1+4\lambda)(1+2\lambda)^2}
\,\,\,.
\label{eq:37xb}
\end{equation}
In Fig.3 we compare different approximations for $\text{Var}
(T_L)$ with numerical data. It is clear that
Eqs.(\ref{eq:35xb}-\ref{eq:37xb}) give a very good agreement. Note
that the region of validity of the standard expression (\ref{var})
obtained in the quadratic approximation is very narrow because the
numerical coefficients at higher terms ($\lambda^3,\,\lambda^4$,
etc) in the expansion of $ {\text {Var}}(T_L)$ grow rapidly.
\begin{figure}[!htb]
\begin{center}
\leavevmode
\epsfysize=7cm
\epsfbox{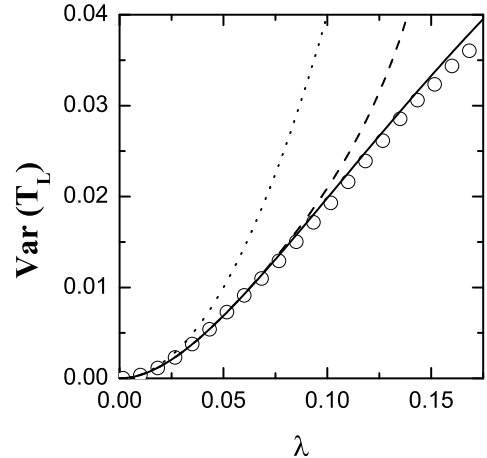}
\narrowtext
\caption{Analytical estimates of the mean variance $\text{Var} (T_L)$
plotted against the numerical data (open circles) for $E=1.5$ and
$W=0.1$, with the average over $10^4$ realizations of disorder.
Dots stand for the standard quadratic approximation (\ref{var}),
dashed lines represent the estimate that takes into account terms
up to the sixth power of $\lambda$ in the expansions of $\left<
\mbox{ln} Z_{L} \right>$ and $\left< \mbox{ln}^2 Z_{L} \right>$,
namely, $\text{Var} (T_L) = 4\lambda^2 - 32\lambda^3 + 176
\lambda^4 - 832
\lambda^5 + 3648\lambda^6$. The full curve corresponds to our
expression (\ref{eq:37xb}). }
\label{fig:5}
\end{center}
\end{figure}

\section{Localized regime.}

In sufficiently long samples a regime of strong localization,
$\lambda =L/l_\infty \gg 1$, is realized. Unlike the previous
case, now all the trajectories of the classical map (\ref{eq:3})
move off the origin of the phase space very fast. This means that
the radii $r_{L, i=1,2}$ are rapidly growing functions of discrete
time $L$. Therefore, in this case $Z_L \gg 1$ and it is convenient
to represent the transmission coefficient $T_L$ in the form,
\begin{equation}
T_{L} = \frac{2}{1+Z_{L}} =
\frac{1}{Z_{L}}\frac{2}{1+\frac{1}{Z_{L}}} = \eta_{L}
\frac{2}{1+\eta_{L}}.
\label{eq:38x}
\end{equation}
Here we introduced a small parameter $\eta_{L} = 1/Z_{L} \ll 1$ in
order to develop a perturbative approach. In the lowest
approximation we have,
\begin{equation}
T_{L} = 2 \, \eta_{L} = \frac{2}{Z_{L}} \, \, .
\label{eq:39x}
\end{equation}
It is known that in a strongly localized regime the transmission
coefficient is exponentially small and reveals very strong
fluctuations. For this reason, $T_L$ is not a self-averaged
quantity, and it is worth to study the logarithm of $T_L$. Its
mean value is given by
\begin{equation}
\left< \mbox{ln}\,T_{L} \right> = \mbox{ln} \, 2
- \left< \mbox{ln}Z_{L} \right>,
\label{eq:40x}
\end{equation}
where the mean value of $\ln Z_L$ again can be expressed in terms
of the classical trajectories of the map (\ref{eq:3}),
\begin{equation}
\left< \mbox{ln}Z_{L} \right> = 2 \lambda - \ln 2 +
R_\infty.
\label{eq:41x}
\end{equation}
Here the effect of correlations between the phases
$\theta_{n}^{(1)}$ and $\theta_{n}^{(2)}$ enters in the last term,
\begin{equation}
\label{term}
R_\infty =
\left< \mbox{ln}
{\left\{ 1 + \exp{ \left [\sum_{n=0}^{L-1} A_{n}
\left( \sin{  2\theta_{n}^{(2)}}
- \sin {2\theta_{n}^{(1)}} \right) \right ]}
\right\} } \right> .
\end{equation}

The detailed analysis \cite{long} of this expression shows that,
in fact, the term $R_\infty$ is independent of the sample length
$L$. To justify this we need to take into account that in the
limit $\lambda_n\rightarrow \infty$ the correlator (\ref{R2})
shown in Fig.1 approaches $1/2$,  that is the mean value of
$\sin^2\theta$. Therefore, the phases $\theta_{0}^{(1)}$ and
$\theta_{0}^{(2)}$ fluctuate coherently in such a way that
\begin{equation}
\sin \left( 2 \theta_{n}^{(2)} \right) -
\sin \left( 2 \theta_{n}^{(1)} \right) \rightarrow  0\,\,\,
\label{eq:25}
\end{equation}
for $n\rightarrow \infty$ (this result is analytically proved in
\cite{long}).

For this reason the upper limit in the sum in Eq.(\ref{term}) can
be replaced by $L =\infty$. Then the correlation term $R_\infty$
becomes $L-$independent. The latter indicates that this term is
$l_\infty$-independent, as well, due to the scaling dependence of
$T_L$ on the parameter $\lambda$. Being $\lambda$ independent, the
correlation term $R_\infty$ is a constant. Unfortunately, we are
unable to evaluate this term analytically. The main difficulty is
that the strength of the correlations between phases changes along
the trajectory, see Fig.1. At the initial stage of evolution when
the correlator (\ref{R2}) is different from $1/2$, the two
trajectories are neither statistically independent nor coherent.
This initial stage gives a contribution, which being small in the
localized regime, nevertheless clearly shows that the statistics
of $\ln Z_L$ and $\ln{T_L}$ is not exactly Gaussian (see below).

We evaluated the term $R_\infty$ numerically and obtained that the
following relation holds with a high accuracy,
\begin{equation}
\label{2ln2}
R_\infty - 2\ln2 =0\,\,\,.
\end{equation}
Substituting Eq.(\ref{2ln2}) into Eq.(\ref{eq:40x}) we get the
standard expression for the mean value of $\ln T_L$,
\begin{equation}
\left< \mbox{ln}\,T_{L} \right> = -2 \lambda \,.
\label{eq:43}
\end{equation}
Since this formula takes into account the correlations along the
whole trajectory, it is accurate up to the zero order term with
respect to $\lambda$.

In the localized regime the transmission coefficient and its
logarithm exhibit strong fluctuations. The approximate
distribution function for $\ln {Z_L}$ can be easily obtained if we
neglect the initial stage of the trajectory and substitute
$r_n^{(1)} \approx r_n^{(2)} $ in Eq.(\ref{eq:4}). Then, for $\ln
{Z_L}$ we obtain
\begin{equation}
\label{lnZ}
\ln Z_L  \approx  r_L^2 =
\sum\limits_{n=1}^{L} \ln D_{n}^2 \approx \sum\limits_{n=1}^{L}
A_n \sin 2\theta_n + \Gamma_2
\,\,\,,
\end{equation}
where
\begin{equation}\label{Sigma2}
\Gamma_2 = \sum\limits_{n=1}^{L}
\left(- \frac{1}{2} A_n^2 \sin^2 2\theta_n
+ A_n^2 \sin^2 \theta_n \right)\,\,\,.
\end{equation}
The first term in the last form of Eq.(\ref{lnZ}) is a sum of
$L>>1$ random independent numbers. Therefore, the statistical
distribution of $\ln Z_L$ is the Gaussian if the quadratic terms
$\Gamma_2$ are neglected. Thus, we can conclude that the log-norm
distribution for the transmission coefficient in the localized
regime is obtained in the lowest approximation with weak disorder,
and neglecting the difference (\ref{eq:25}) between phases along
the two trajectories.

The parameters of the  Gaussian distribution for $\ln Z_L $ are
calculated from Eq.(\ref{lnZ}),  where the first (linear) term has
the zero mean, however, a wide dispersion. The non-zero correction
comes from the second (quadratic) term, $\left<
\ln Z_L \right> = \left< \Sigma_2 \right> \approx
\frac{1}{4} L \left<A_n^2\right> = 2 \lambda$.
Calculating the second moment of $\ln Z_L$ we can neglect the
quadratic term in Eq.(\ref{lnZ}) and get,
\begin{equation}\label{lnZ2}
\left< \ln^2 Z_L \right> \approx 4\lambda + (2\lambda)^2\,\,\,.
\end{equation}
Then, for the variance of $\ln Z_L$ we have,
\begin{equation}\label{sigma2}
\sigma^2 \equiv \left< \ln^2 Z_L \right> -  \left< \ln Z_L
\right> ^2 \approx 4\lambda \, ,
\end{equation}

Now we can write the distribution function for $\ln Z_L$,
\begin{equation}
{\cal P}\left(\mbox{ln}Z_{L}\right)  =
\frac{1}{\sqrt{2\pi\sigma^{2}}}
\, \exp\left[-\frac{\left(\mbox{ln}Z_{L}-\left<\mbox{ln}Z_L\right>
\right)^{2}}{2\sigma^{2}}
\right]\,\,\, .
\label{eq:pdlnZ_L}
\end{equation}

In Fig.4 we fit the numerical data obtained for $\lambda =10$ by
the Gaussian distribution Eq.(\ref{eq:pdlnZ_L}), using the
dispersion $\sigma$ as a fitting parameter. The best fit was
obtained for $\sigma^2=4\lambda + C_0$ with $C_0 =3.2$, and
$\left<\mbox{ln}Z_L\right>=2\lambda + \ln 2$. The small correction
$C_0$ to the dispersion $4\lambda$ originates from the initial
stage of the trajectories that is neglected in Eq.(\ref{sigma2}).
At the same time, the center of the distribution (the mean value)
is in agreement with Eq.(\ref{eq:41x}) where the calculation is
performed exactly.
\begin{figure}[!htb]
\begin{center}
\leavevmode
\epsfysize=8cm
\epsfbox{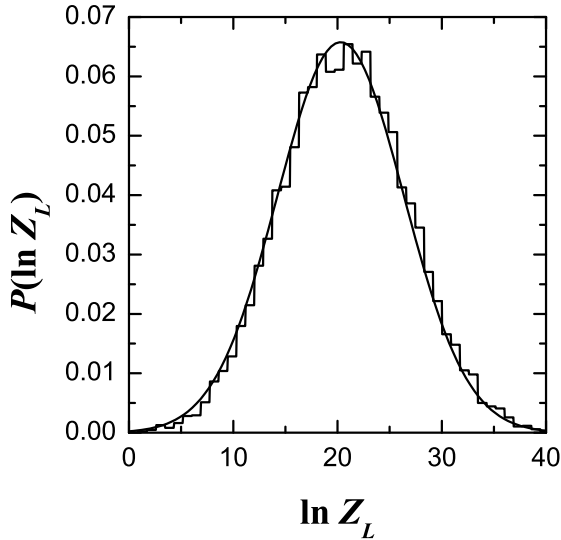}
\caption{\small Numerical data (broken curve) for the probability
distribution of $\mbox{ln}Z_{L}$ plotted against
\mbox{Eq.(\ref{eq:pdlnZ_L})} (smooth curve). Numerical data were
obtained for $E=1.5$, $W=0.1$ and $\lambda=L/l_{\infty}=10$ for
$10^4$ realizations of the disorder.}
\label{fig:pdlnZ_L}
\narrowtext
\end{center}
\end{figure}

It should be noted that if the parameter $\lambda$ is not very
big, the numerical histogram for ${\cal P} (\ln Z_L)$ manifests an
asymmetry. The left tail reveals a natural cutoff at $\ln Z_L = 0$
since $\ln{Z_L}$ is a positive function, see Eqs.(\ref{eq:6}) and
(\ref{eq:8}). This asymmetry is not visible as long as the
dispersion $\sigma = 2\sqrt{\lambda}$ is much less than the mean
value $2\lambda$.

The log-norm distribution Eq.(\ref{eq:pdlnZ_L}) leads to the
following mean value for the transmission coefficient \cite{long},
\begin{equation}
\left<T_{L}\right>
\approx \left(\frac{\pi}{2}\right)^{\frac{1}{2}}
\left(\frac{L}{l_{\infty}}\right)^{-\frac{1}{2}}
\exp\left(-\frac{L}{2l_{\infty}}\right)\,\,\,.
\label{approx}
\end{equation}
This result is slightly different from the exact formula (see,
e.g, \cite{LGP88,Makarov}),
\begin{equation}
\left<T_{L}\right> \approx
\left( \frac{\pi}{2}\right)^{\frac {5}{2}}
\left(\frac{L}{l_{\infty}}\right)^{-\frac{3}{2}}
\exp\left(-\frac{L}{2l_{\infty}}\right)\,\,\,.
\label{exact}
\end{equation}
Having the same exponential dependence, the approximate expression
(\ref{approx}) differs from the exact one by an extra factor
$(\pi/2)^2\left(L/l_\infty \right)^{-1}$. This discrepancy
originates from the above-mentioned contribution of the initial
stage of evolution that is neglected in the distribution function
Eq.(\ref{eq:pdlnZ_L}).

\section{Conclusions.}

We have studied the transport properties of the 1D standard
Anderson model with weak random potential, using the Hamiltonian
map approach. This approach is based on a reduction of the quantum
model to the classical two-dimensional map that describes the
dynamics of a linear parametric oscillator with a delta-kick time
dependence of its frequency. Amplitudes of the kicks are
determined by the site potential of the original quantum model,
and the energy of an eigenstate enters into the map as a free
parameter.

Some results have been already obtained with the use of this
approach in application to both uncorrelated and correlated random
potentials \cite{IK99,KI00,IKU01,IKT95,KTI97,IRT98,Luca}. All
these studies refer to the properties of the localization length
in {\it infinite} samples. In contrast to previous results, in
this paper we consider a new question about the effectiveness of
the Hamiltonian map approach in application to transport
properties of {\it finite} samples. Specifically, we are
interested in the mean values of the transmission coefficient
$T_L$, its second moment, variance, and the distribution
functions.

We performed an analytical treatment for two limit cases of
ballistic and strongly localized regimes assuming weakness of the
random potential. For the ballistic regime (when $l_\infty\gg L$)
we were able to derive analytical expressions for the mean values
$\left< T_L \right>\, , \left< T_L^2 \right> $ and the variance
$\left< T_L^2\right> - \left< T_L \right>^2$, that are much more
accurate that the standard estimates known in the literature. The
analysis has revealed a non-trivial role of the correlations
between two complimentary classical trajectories that determine
the transmission coefficient. Our numerical study confirms the
analytical predictions.

For strongly localized regime, $l_\infty \ll L$, our main interest
was in the mean value of the logarithm of $T_L$ and in the
distribution function for $\ln T_L$. We have found that the
leading term for $\left< \ln T_L \right>$ can be easily obtained
if one neglects the initial stage of the evolution of the
complementary classical trajectories. It is also easy to show that
the distribution of $\ln T_L$ has the Gaussian form. In terms of
the classical map, this log-norm distribution results from the
central limit theorem applied to the expression for the radius of
classical trajectories. Correspondingly, the approach easily
reproduces the estimates for the mean value of the second moment
of $\ln T_L$ and for its variance.

The results of our analysis may find further applications to the
problem of the single parameter scaling, as well as in the study
of transport properties of 1D random models with long-range
correlations in the site potential.

\section {Acknowledgments}
The authors are very thankful to Luca Tessieri for his valuable
comments and fruitful discussions. This research was supported by
Consejo Nacional de Ciencia y Tecnolog\'{\i}a (CONACYT, M\'exico)
grant 34668-E.

\end{multicols}

\end{document}